Scientific
Research
Publishing

# The Dynamics of Wave-Particle Duality

## Adriano Orefice*, Raffaele Giovanelli, Domenico Ditto

Department of Agricultural and Environmental Sciences (DISAA), University of Milan, Milano, Italy
Email: *adriano.orefice@unimi.it

## Abstract

Both classical and wave-mechanical monochromatic waves may be treated in terms of *exact* ray-trajectories (*encoded in the structure itself of Helmholtz-like equations*) whose mutual coupling is the one and only cause of any diffraction and interference process. In the case of Wave Mechanics, de Broglie's merging of Maupertuis's and Fermat's principles (see Section 3) provides, *without resorting to the probability-based guidance-laws and flow-lines of the Bohmian theory*, the simple law addressing particles along the *Helmholtz rays* of the relevant matter waves. The purpose of the present research was to derive the *exact* Hamiltonian ray-trajectory systems concerning, respectively, *classical electromagnetic waves*, *non-relativistic matter waves* and *relativistic matter waves*. We faced then, as a typical example, the numerical solution of non-relativistic wave-mechanical equation systems in a number of numerical applications, showing that each particle turns out to "dances a *wave-mechanical* dance" around its *classical* trajectory, to which it reduces when the ray-coupling is neglected. Our approach reaches the double goal of a clear insight into the mechanism of wave-particle duality and of a reasonably simple computability. We finally compared our exact *dynamical* approach, running as close as possible to Classical Mechanics, with the *hydrodynamic* Bohmian theory, based on fluid-like "guidance laws".

## 1. Introduction

As is well expressed in Ref. [1], "*the knowledge of several routes and their connections is always helpful when traveling through the quantum territory*". The authors of Ref. [1] had in mind, however, a "territory" crossed by routes decla-redly leading to Bohm's Mechanics [2]-[7], with its probability-based guidance-laws and *hydrodynamic flow-lines*.

Having in mind, on our part, a set of *exact particle trajectories running as close as possible to the ones of Classical Dynamics,* the route adopted in the present paper starts from the very foundations of Wave Mechanics [8] [9], *i.e.* from

1) de Broglie's founding principle $p = \hbar k$, to which we owe the very concept of "wave-particle duality", associating for the first time a "matter wave" (of wave-vector $k$) with a particle of momentum $p$, being $\hbar = h/2\pi$ and $h$ the Planck constant [10] [11];

2) Schrödinger's *time-independent* equation [12] [13] [14] [15], bypassing (with its eigen-fuctions and eigen-values) the heuristic prescriptions of the "old" Quantum Mechanics: an equation whose *energy-dependence* suggests a crucial analogy with Classical Mechanics, to be exploited in the present paper;

3) Schrödinger's *time-dependent* equation, whose *energy-independence* suggests a probabilistic distribution over the entire eigen-energy spectrum; and finally

4) Davisson-Germer's diffraction test of the real existence of de Broglie's matter waves in physical space [16].

We demonstrate, to begin with, that *any Helmholtz-like equation* is associated with a characteristic Hamiltonian set of "rays" in physical space, allowing an *exact*, *trajectory-based* approach to any kind of *classical* wave-like process, bypassing any geometrical optics approximation.

Passing then to the *wave-mechanical* case, de Broglie's relation $p = \hbar k$ clearly tells us that the *Helmholtz rays* (with wave vector $k$) of de Broglie's matter waves coincide with the *dynamical trajectories* of the associated particles with momentum $p$, thus laying *the foundations of Wave Mechanics*. Each particle is guided along its path by the energy-preserving "gentle drive" of a matter wave having a well-established physical existence [16]. Our approach reaches the double goal of a clear insight into the mechanism of wave-particle duality and of a simple numerical computability: a necessary condition for any theory of practical interest.

An *experimentally tested* Hamiltonian description of wave-like features holding beyond the eikonal approximation [17] was first contrived in Refs. [18] [19] [20] for the propagation of electromagnetic waves in thermonuclear plasmas. This method was applied until recently [21] to the microwave diagnostics installed on the TORE SUPRA fusion facility, and extended in Ref. [22] to the treatment of the Schrödinger equation. The general demonstration of the equivalence between classical and wave-mechanical Helmholtz-like equations and exact, trajectory-based Hamiltonian systems was given in Refs. [23] [24] [25] [26] [27], and led to extensive application both in relativistic electrodynamics and in the analysis of experimental arrangements and devices employed for light transmission and guiding [28] [29] [30]. Let us finally remind that the equivalence between the time-dependent Schrödinger equation and Hamiltonian Me-chanics was demonstrated in Ref. [31] in terms of symplectic geometry, without giving rise, so far, to numerical applications.

We derive here the *exact* Hamiltonian systems concerning, respectively, *classical* electromagnetic waves (Section 2), *non-relativistic* matter waves (Section 3) and *relativistic* matter waves (Section 4). Examples of numerical computation are given in Section 5, and a comparison is made in Sections 6 and 7 with the hydrodynamic approach of the Bohmian theory.

## 2. The Case of Classical Waves

We shall refer in the present Section, in order to fix ideas, to a *stationary*, isotropic and (generally) inhomogeneous dielectric medium, sustaining a classical *monochromatic* electromagnetic wave whose electric and/or magnetic field components are represented in the form

$$\psi(\boldsymbol{r}, \omega, t) = u(\boldsymbol{r}, \omega) e^{-i\omega t}, \tag{1}$$

where $\boldsymbol{r} \equiv (x, y, z)$ and $u(\boldsymbol{r}, \omega)$ is assumed to satisfy the Helmholtz equation

$$\nabla^2 u + (nk_0)^2 u = 0, \tag{2}$$

with

$$\underline{\nabla} \equiv \partial/\partial \boldsymbol{r} \equiv (\partial/\partial x, \partial/\partial y, \partial/\partial z); \quad \nabla^2 = \frac{\partial^2}{\partial x^2} + \frac{\partial^2}{\partial y^2} + \frac{\partial^2}{\partial z^2}; \quad k_0 \equiv \frac{2\pi}{\lambda_0} = \frac{\omega}{c} \tag{3}$$

and with a (time-independent) refractive index $n \equiv n(\boldsymbol{r}, \omega)$. The time-independence of Equation (2) does NOT mean, of course, that no physical change is expected. Just as for the usual laws of Dynamics, once suitable boundary and initial conditions are assigned, the ensuing motion, occurring in *stationary external conditions*, is exactly described. If we perform in fact, into Equation (2), the quite general replacement

$$u(\boldsymbol{r}, \omega) = R(\boldsymbol{r}, \omega) e^{i\varphi(\boldsymbol{r}, \omega)}, \tag{4}$$

with real amplitude $R(\boldsymbol{r}, \omega)$ and real phase $\varphi(\boldsymbol{r}, \omega)$, and separate real from imaginary parts, Equation (2) is seen to provide [23] a *kinematical* "ray-trajectory" system of the Hamiltonian form

$$\begin{cases} \dfrac{d\boldsymbol{r}}{dt} = \dfrac{\partial D}{\partial \boldsymbol{k}} = \dfrac{c\boldsymbol{k}}{k_0} \\[2mm] \dfrac{d\boldsymbol{k}}{dt} = -\dfrac{\partial D}{\partial \vec{r}} = \boldsymbol{\nabla}\left[ \dfrac{ck_0}{2} n^2(\boldsymbol{r}, \omega) - W(\boldsymbol{r}, \omega) \right] \\[2mm] \boldsymbol{\nabla} \cdot (R^2 \boldsymbol{k}) = 0 \\[2mm] k(t = 0) \equiv k_0 = \omega/c \end{cases} \tag{5}$$

where

$$\boldsymbol{k} = \boldsymbol{\nabla}\varphi, \tag{6}$$

$$D(\boldsymbol{r}, \boldsymbol{k}, \omega) \equiv \frac{c}{2k_0}\left[ k^2 - (nk_0)^2 - \frac{\nabla^2 R}{R} \right], \tag{7}$$

$$W(\boldsymbol{r}, \omega) \equiv -\frac{c}{2k_0} \frac{\nabla^2 R(\boldsymbol{r}, \omega)}{R(\boldsymbol{r}, \omega)}, \tag{8}$$

and a "ray" velocity $v_{ray} = ck/k_0$ is implicitly defined. It is easily seen that, as long as $k \equiv |\mathbf{k}| = k_0$, we'll have $v_{ray} \equiv |\mathbf{v}_{ray}| = c$. The function $W(\mathbf{r}, \omega)$ (which we call "Wave Potential"), represents a *newly discovered* intrinsic property encoded *in any Helmholtz-like equation*. It is seen to couple together the whole set of ray-trajectories, causing (thanks to its frequency dependence) any diffraction and/or interference process. Its gradient $\nabla W$ turns out to be orthogonal to $\mathbf{k} \equiv \nabla \varphi$, thus modifying the direction, but not the amplitude, of the wave vector $\mathbf{k}$ itself. The time-integration of the Hamiltonian system (5) requires the full knowledge of the amplitude $R(\mathbf{r}, \omega)$ on an assigned starting wave-front, together with the ray positions $\mathbf{r}(t = 0)$ and of the corresponding wave vectors $\mathbf{k}(t = 0)$, orthogonal to the wave-front. Thanks to the third of Equations (5) (expressing the constancy of the flux of $R^2 \nabla \varphi$ along any tube formed by the field lines of the wave vector field $\mathbf{k} = \nabla \varphi$, *i.e.* by the "ray-trajectories" themselves) one obtains, then, the amplitude $R(\mathbf{r}, \omega)$ over the next wave-front. The iteration of this procedure allows to build up, step by step, *both* the "Helmholtz trajectories" along which the "rays" are channeled *and* the time-table of the "ray motion" along them. We shall see at the end of Section 4 what the "rays" are conveying.

When, in particular, the space variation length $L$ of the amplitude $R(\mathbf{r}, \omega)$ satisfies the condition $k_0 L \gg 1$, the Wave Potential term $W(\mathbf{r}, \omega)$ may be dropped, thus removing any ray-coupling. The "rays" will therefore propagate independently of each other, without any diffraction and/or interference, according to the *eikonal equation* [17]

$$k^2 \cong (nk_0)^2. \tag{9}$$

## 3. The Case of Non-Relativistic de Broglie's Matter Waves

Let us pass now to the case of non-interacting particles of mass $m$ and total energy $E$, launched with an initial momentum $\mathbf{p}_0$ (with $p_0 = \sqrt{2mE}$) into an external force field deriving from a *time-independent* potential energy $V(\mathbf{r})$. The *classical* dynamical behavior of each particle is described, as is well known [17], by the time-independent Hamilton-Jacobi (H-J) equation

$$(\nabla S)^2 = 2m[E - V(\mathbf{r})], \tag{10}$$

where the basic property of the H-J function $S(\mathbf{r}, E)$ is that the particle momentum is given by

$$\mathbf{p} = \nabla S(\mathbf{r}, E). \tag{11}$$

The *classical* H-J surfaces $S(\mathbf{r}, E) = const$, perpendicular to the momentum of the moving particles, *pilot* them along stationary trajectories, according to the laws of Classical Mechanics.

Let us remind now [17] that, according to *Fermat's* variational principle, any *ray* of a wave travelling (with wave-vector $\mathbf{k}$) between two points $A$ and $B$ shall follow a trajectory satisfying the condition $\delta \int_A^B k \, \mathrm{d}s = 0$, where $k = |\mathbf{k}|$ and $\mathrm{d}s$ is an element of a virtual line connecting $A$ and $B$. According to *Maupertuis's* variational principle, on the other hand, any *material particle* moving between $A$ and $B$ with momentum $\boldsymbol{p}$ shall follow a trajectory satisfying the analogous condition $\delta \int_A^B p \, ds \equiv 0$, with $p = |\boldsymbol{p}|$. This striking analogy between the mechanical behavior of waves and matter induced Louis de Broglie to associate each *moving particle* with a suitable "*matter wave*" of the form, suggested by Equations (1) and (4),

$$\psi = u(\boldsymbol{r}, \omega) e^{-i\omega t} \equiv R(\boldsymbol{r}, \omega) e^{i[\varphi(\boldsymbol{r}, \omega) - \omega t]} \tag{12}$$

under Planck's condition

$$\omega = E/\hbar , \tag{13}$$

according to the basic conjecture

$$\boldsymbol{p} = \hbar \boldsymbol{k} \tag{14}$$

laying *the very foundations of Wave Mechanics* [10] [11]. We have therefore, thanks to Equations (6) and (11)-(14), the relations $\varphi = S(\boldsymbol{r}, E)/\hbar$ and

$$u(\boldsymbol{r}, E) \equiv R(\boldsymbol{r}, E) e^{\frac{i}{\hbar} S(\boldsymbol{r}, E)} , \tag{15}$$

showing that the H-J surfaces $S(\boldsymbol{r}, E) = const$ represent the phase-fronts of matter waves, while maintaining the *piloting* role played in the classical Equation (11). Equation (14) *provides both the structure* (15) *and the "guidance equation" of de Broglie's matter waves, addressing the particles with momentum $\boldsymbol{p}$ along the wave-vector $\boldsymbol{k}$*. Point-particles are driven, in other words, along stationary trajectories orthogonal to the phase-fronts of matter waves with $\lambda \equiv \frac{2\pi}{k} = 2\pi\hbar/p$. The successive step was performed by Schrödinger ([12] [13] [14] [15]), with the suggestion of *viewing the laws of Classical Mechanics*—represented here by Equation (10)—*as the eikonal approximation of a suitable Helmholtz-like equation*. Reminding that Equation (9) provides the eikonal limit of the Helmholtz Equation (2), this suggestion simply amounts to performing, into Equation (2), making use of Equations (9)-(14), the replacement,

$$(n k_0)^2 \rightarrow k^2 \equiv \frac{p^2}{\hbar^2} = \frac{2m}{\hbar^2}(E - V) \tag{16}$$

leading to the Helmholtz-like equation

$$\nabla^2 u(\boldsymbol{r}, E) + \frac{2m}{\hbar^2}\big[E - V(\boldsymbol{r})\big] u(\boldsymbol{r}, E) = 0 . \tag{17}$$

which is the well-known *time-independent* (and *energy-dependent*) *Schrödinger equation*, holding for stationary matter waves. It's an eigen-value equation admitting in general both continuous and discrete energy and eigen-function spectra, which replace the heuristic prescriptions of the "old" quantum theory [8].

The *real existence of de Broglie's matter waves in physical space* was established in 1927 by the Davisson-Germer experiments [16] on electron diffraction by a nickel crystal.

Having in mind the dynamics of particles with an *assigned total energy, just as it usually occurs in classical physics*, let us now apply to the Helmholtz-like *energy-dependent* Equation (17) the same procedure leading from the Helmholtz Equation (2) to the Hamiltonian ray-tracing system (5), by simply replacing Equation (15) into Equation (17) and separating real and imaginary parts. After having defined the energy-dependent "Wave Potential function"

$$W(\boldsymbol{r}, E) = -\frac{\hbar^2}{2m} \frac{\nabla^2 R(\boldsymbol{r}, E)}{R(\boldsymbol{r}, E)}.$$  (18)

and the energy function

$$H(\boldsymbol{r}, \boldsymbol{p}, E) \equiv \frac{p^2}{2m} + W(\boldsymbol{r}, E) + V(\boldsymbol{r})$$  (19)

we get now the *dynamical* Hamiltonian system

$$\begin{cases} \dfrac{\mathrm{d}\boldsymbol{r}}{\mathrm{d}t} = \dfrac{\partial H}{\partial \boldsymbol{p}} \equiv \dfrac{\boldsymbol{p}}{m} \\[2mm] \dfrac{\mathrm{d}\boldsymbol{p}}{\mathrm{d}t} = -\dfrac{\partial H}{\partial \boldsymbol{r}} \equiv -\boldsymbol{\nabla}\big[V(\boldsymbol{r}) + W(\boldsymbol{r}, E)\big] \\[2mm] \boldsymbol{\nabla} \cdot (R^2 \boldsymbol{p}) = 0 \\[2mm] p(t=0) \equiv p_0 = \sqrt{2mE} \end{cases}$$  (20)

providing, in strict analogy with the Helmholtz ray-tracing system (5), a stationary set of particle trajectories, *together with the time-table of the particle motion along them*. These trajectories are mutually coupled, once more, by the "*Wave Potential*" function $W(\boldsymbol{r}, E)$ of Equation (18), acting orthogonally to the particle motion and exerting an energy-conserving "gentle drive" which is the cause, thanks to its "monochromaticity", of any diffraction and/or interference process. Notice that this *energy-dependence* makes $W(\boldsymbol{r}, E)$ basically different (in spite of its formal analogy) from Bohm's *energy-independent* "Quantum Potential" $Q(\boldsymbol{r}, t) = -\dfrac{\hbar^2}{2m} \dfrac{\nabla^2 \psi(\boldsymbol{r}, t)}{\psi(\boldsymbol{r}, t)}$ [1]-[7]. The third of Equations (20) expresses the constancy of the flux of $R^2 \nabla S$ along any tube formed by the field lines of $\boldsymbol{p} = \nabla S(\boldsymbol{r}, E)$, *i.e.* by the trajectories themselves. In complete analogy with Section 2 the launching values $\boldsymbol{r}(t=0)$ and $\boldsymbol{p}(t=0)$ of each particle must be supplemented, in the time-integration of the system (20), with the assignment of the amplitude $R(\boldsymbol{r}, E)$ (and its derivatives) over a starting wave-front, orthogonal to $\boldsymbol{p}(t=0)$ at each point $\boldsymbol{r}(t=0)$. The third of equations (20) provides then, step by step, the function $R(\boldsymbol{r}, E)$ (with its derivatives) over the next wave-front, and so on. Notice that the Wave Potential (18) depends on the profile of the transverse particle distribution $R(\boldsymbol{r}, E)$, but not on its intensity, which may be quite arbitrarily chosen.

## 4. The Case of Relativistic de Broglie's Matter Waves

Let us finally pass (maintaining the same notations of the previous Sections) to the relativistic dynamics of particles with rest mass $m_0$ and assigned energy $E$, launched, as in the previous case, into a force field deriving from a stationary potential energy $V(r)$, and moving according to the (relativistic) *time-independent* Hamilton-Jacobi equation [8] [9]

$$\left[\nabla S(r,E)\right]^2 = \left[\frac{E-V(r)}{c}\right]^2 - (m_0 c)^2 . \tag{21}$$

After having repeated de Broglie's logical steps (11)-(14), we shall assume, with Schrödinger, that the relevant matter waves satisfy a *Helmholtz-like* equation of the form (2), reducing to *standard Mechanics*—represented now by Equation (21)—in its *eikonal approximation*, and perform therefore, into Equation (2), the replacement

$$(nk_0)^2 \to k^2 \equiv \frac{p^2}{\hbar^2} = \left[\frac{E-V(r)}{\hbar c}\right]^2 - \left(\frac{m_0 c}{\hbar}\right)^2 \tag{22}$$

thus obtaining the *time-independent* Klein-Gordon equation [8] [9]

$$\nabla^2 u + \left[\left(\frac{E-V(r)}{\hbar c}\right)^2 - \left(\frac{m_0 c}{\hbar}\right)^2\right] u = 0 , \tag{23}$$

holding for de Broglie's relativistic matter waves associated with particles of total energy $E$.

By replacing Equations (15) into Equation (23) and separating once more real and imaginary parts, Equation (23) leads [26] to the dynamical Hamiltonian system

$$\begin{cases} \dfrac{dr}{dt} = \dfrac{\partial H}{\partial p} \equiv \dfrac{c^2 p}{E-V(r)} \\[2mm] \dfrac{dp}{dt} = -\dfrac{\partial H}{\partial r} \equiv -\nabla V(r) - \dfrac{E}{E-V(r)} \nabla W(r,E) \\[2mm] \nabla \cdot (R^2 p) = 0 \\[2mm] p(t=0) \equiv p_0 = \sqrt{(E/c)^2 - (m_0 c)^2} \end{cases} \tag{24}$$

where

$$H(r,p,E) \equiv V(r) + \sqrt{(pc)^2 + (m_0 c^2)^2 - \hbar^2 c^2 \frac{\nabla^2 R(r,E)}{R(r,E)}} \tag{25}$$

and a Wave Potential

$$W(r,E) = -\frac{\hbar^2 c^2}{2E} \frac{\nabla^2 R(r,E)}{R(r,E)} \tag{26}$$

couples and guides, once more, without any wave-particle energy exchange, the particle trajectories, acting *orthogonally* to the particle motion. It is interesting to observe that the first of Equations (24) turns out to coïncide with the "guidance formula" presented by de Broglie in his relativistic "*double solution theory*" [32], and *doesn't coïncide* with $p/m$, while maintaining itself parallel to the momentum **p** . Let us notice that, in the particular case of *massless* particles (*i.e.* for $m_0 = 0$), the Klein-Gordon Equation (23), thanks to Equation (13), reduces to the form

$$\nabla^2 u + \left[\frac{\hbar\omega - V(\mathbf{r})}{\hbar c}\right]^2 = 0 ,\qquad(27)$$

structurally analogous to Equation (2), which may be viewed therefore as a suitable Klein-Gordon equation holding for *massless* particles. Einstein's first historical duality (conceived for the electromagnetic radiation field [33]) is therefore included in de Broglie's wave-particle mechanism.

## 5. Numerical Examples

We compute in the present Section, for a number of different experimental set-ups, the stationary sets of particle trajectories provided by the integration of the Hamiltonian system (20), which is a direct consequence, as we know, of Schrödinger's energy-dependent Equation (17).

The very plausibility of the numerical results plays in favor of the underlying philosophy. No kind of measurement perturbation is taken into account, and the geometry is assumed, for simplicity sake, to allow a computation limited to the *(x, z)*-plane.

Because of the orthogonality between the wave front and the particle momentum we shall make use, over the (*x, z*) plane, of the identities

$$\left(p_x \frac{\partial}{\partial x} + p_z \frac{\partial}{\partial z}\right) R = 0 \qquad(28)$$

$$\nabla^2 R = \left(p/p_z\right)^2 \partial^2 R/\partial x^2 . \qquad(29)$$

We show in **Figure 1** the diffraction of a Gaussian beam of the initial form $R(x; z = 0) \propto \exp\left(-x^2/w_0^2\right)$, launched from the left hand side along the z-axis, in the absence of external fields (*i.e.* for $V(x,z) = 0$), with $p_x(t = 0) = 0$ and $p_z(t = 0) = \sqrt{2mE}$ . The numerical computations are performed by assuming a ratio $\lambda_0/w_0 \cong 10^{-4}$ between the beam wavelength $\lambda_0$ and "waist" $w_0$, and by plotting the figures in terms of the dimensionless coordinates $x/w_0$ and $z/w_0$ . The (axially symmetrical) particle beam travels under the coupling role of the Wave Potential, which induces a "diffractive widening" due to a progressive conversion of $p_z$ into $p_x$, while respecting the overall conservation of the beam energy and momentum.

The two heavy lines in **Figure 1** represent the trajectories starting (at $z/w_0 = 0$) from the so-called "waist" positions $x/w_0 = \pm 1$, *whose numerical values turn out to be in excellent agreement with their well-known "paraxial"* [34] *analytical expression*

$$\frac{x}{w_0} = \pm\sqrt{1 + \left(\frac{\lambda_0 z}{\pi w_0^2}\right)^2} . \qquad(30)$$

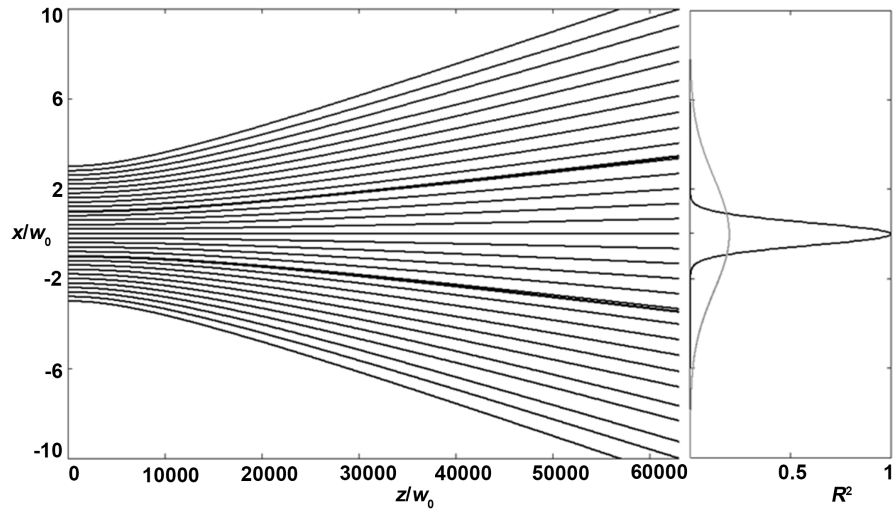

**Figure 1.** Diffraction of a Gaussian beam. The initial and final transverse intensity distributions $R^2$ of the beam are plotted on the *right-hand side*, and the relevant trajectory pattern is plotted on the *left-hand side*.

**Figure 2** refers, in its turn, to the diffraction/interference of two neighbouring Gaussians beams, symmetrical with respect to the axis $x/w_0 = 0$.

**Figure 3** presents, in its turn, the launch, stopping and "backward fall" of a single Gaussian beam of particles with total energy $E$, traveling under a constant force field $-F_{oz}$ acting in the negative $z$-direction. Starting from $z/w_0 = 0$ the beam travels, for a while, in the positive $z$ direction, and is stopped and pushed back, for $z \cong E/F_{oz}$, by the force field, while continuing its diffractive widening due to the Wave Potential.

We shall consider now

1) the case of a *potential barrier* of the form

$$V(z) = V_0 \exp\left[-2(z - z_B)^2 \big/ d^2\right] \tag{31}$$

(where the parameters $z_B$ and $d$ determine the position of the *peak* and of the distance between the flexes, respectively), and

2) the case of a "*step-like*" *potential* of the form

$$V(z) = V_0 \left\{1 + \exp\left[-\alpha \frac{z - z_S}{w_0}\right]\right\}^{-1} \tag{32}$$

where $\alpha$ and $z_S$ determine, respectively, the slope and the *flex* position of the continuous line connecting the two asymptotic levels where $V(z \to -\infty) = 0$ and $V(z \to \infty) = V_0$.

We plot in **Figure 4**, for $z_B/w_0 = z_S/w_0 = 10^4$, the profiles of the respective ratios $V(z)/V_0$, and consider the launch from the left hand side of the Gaussian beam of **Figure 1**, with energy $E$ and waist $w_0$, into these external fields.

In the case of the *potential barrier* (31) the beam gradually widens under the diffractive action of the Wave Potential, and is stopped and thrown back when $E = V(z) < V_0$, with a behavior quite similar to the one of **Figure 3**. When $E/V_0 \gg 1$, on the other hand, the beam overcomes the top of the barrier and undergoes a strong acceleration beyond it. We omit in both cases, for brevity sake, the relevant plots, limiting ourselves to the remarkable case $E/V_0 \cong 1$ (**Figure 5**).

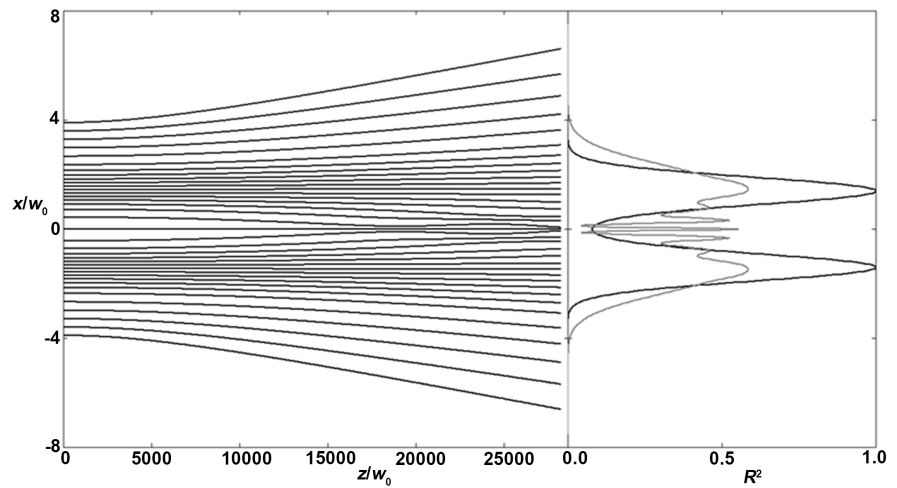

**Figure 2.** The case of two symmetric neighbouring Gaussians beams. The initial and final transverse intensity distributions $R^2$ are plotted on the *right-hand side*, and the relevant trajectory pattern is plotted on the *left-hand side*.

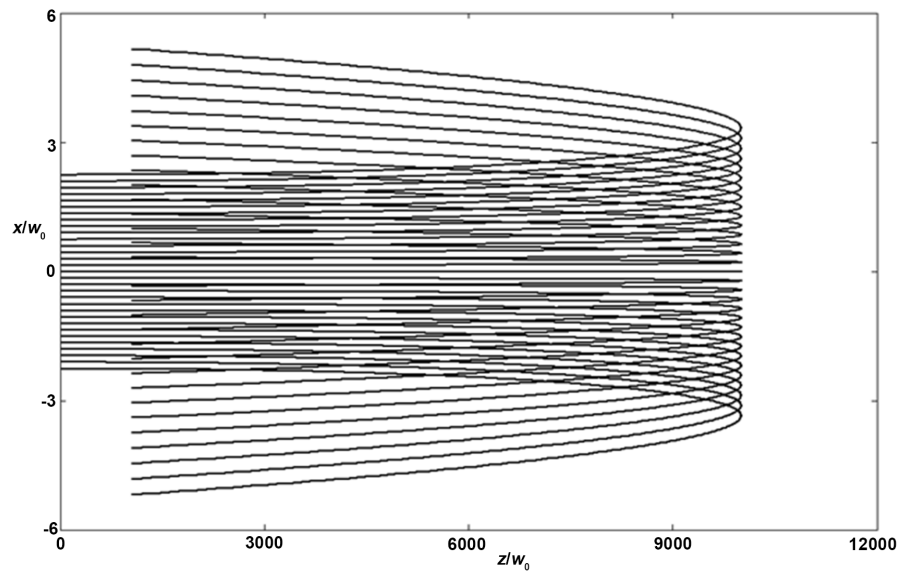

**Figure 3.** Beam launched into a constant field $-F_{oz}$.

Here, in a narrow region around the peak position $z = z_B \equiv 10^4 w_0$, both the external force $F_z(z)$ and $p_z$ are very close to zero. The beam motion basically occurs, under the action of its Wave Potential, along the $x$-axis, and is *evanescent* in the $z$-direction. After this brief interlude, the beam is strongly accelerated along $z$ for $z > z_B$.

In the case of the *step-like potential* (32), the beam gradually widens under the diffractive action of the Wave Potential, and is stopped and thrown back, once more, for $E = V(z) < V_0$, with a behavior similar to the one of **Figure 3**. When $E/V_0 > 1$, on the other hand, the beam travels forward, under the diffractive widening action of the Wave Potential, with an energy reduced by the overcoming of the potential step. We omit, in both cases, the relevant plots, limiting once more ourselves to the remarkable case $E/V_0 \cong 1$ (**Figure 6**). When $V(z)$ is close to $V_0$ both the external force $F_z(z)$ and $p_z$ are close to zero. The beam basically moves, under the action of the Wave Potential alone, along the x-axis, while its forward motion becomes, *and remains*, evanescent.

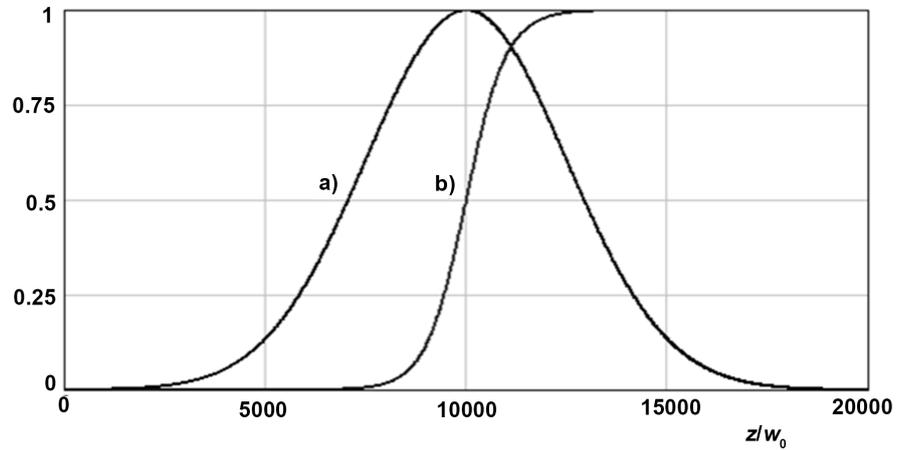

**Figure 4.** Profiles $V(z)/V_0$ of (a) the barrier and (b) the step-like potentials.

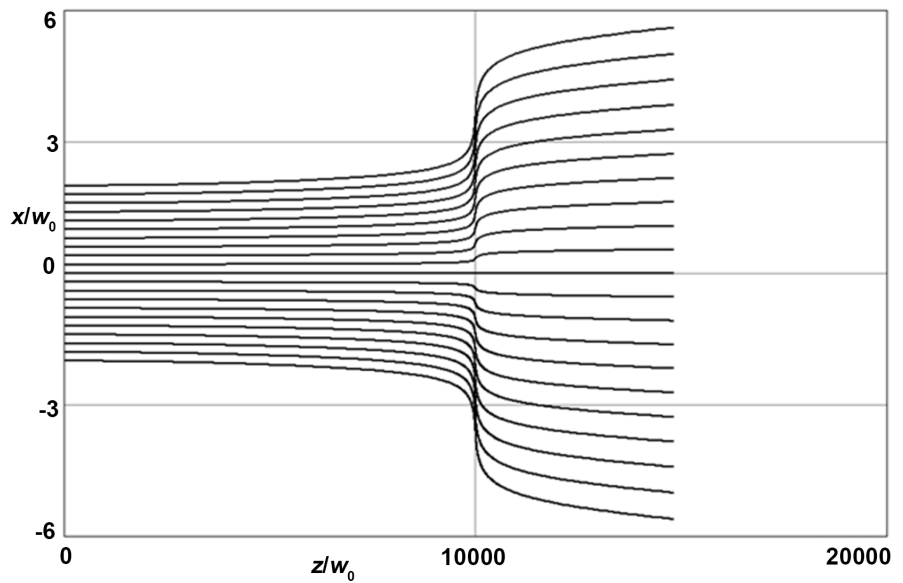

**Figure 5.** Potential barrier: case $E/V_0 \cong 1$.

As a further applicative example, let us come to a collimated matter wave beam launched into a potential field $V(x,z)$ representing a *lens-like* focalizing structure, of which we omit here, for simplicity sake, the analytical expression (see, for instance, Refs. [18] and [22]). **Figure 7** and **Figure 8** present, in this case, the results obtained by neglecting, respectively, and by taking into account, in Equations (25), the Wave Potential term $W(\mathbf{r}, E)$. Its diffractive effect is seen to replace the *point-like* focus of geometrical optics by a *finite focal waist*, strictly reminding the case of the Luneburg lenses considered in Refs. [29] [30].

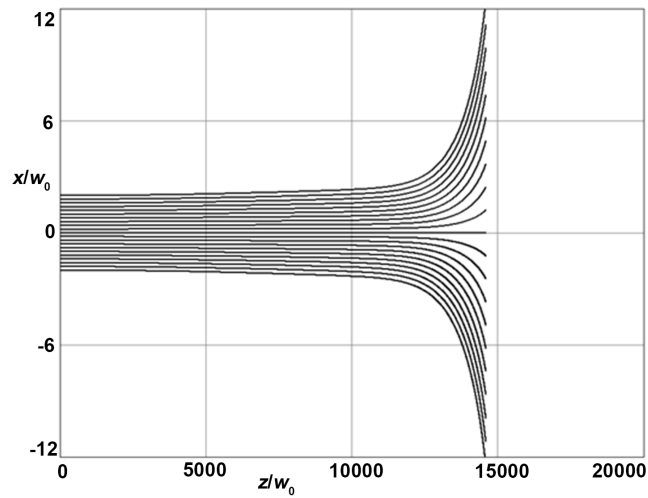

**Figure 6.** Step-like potential: case $E/V_0 \cong 1$.

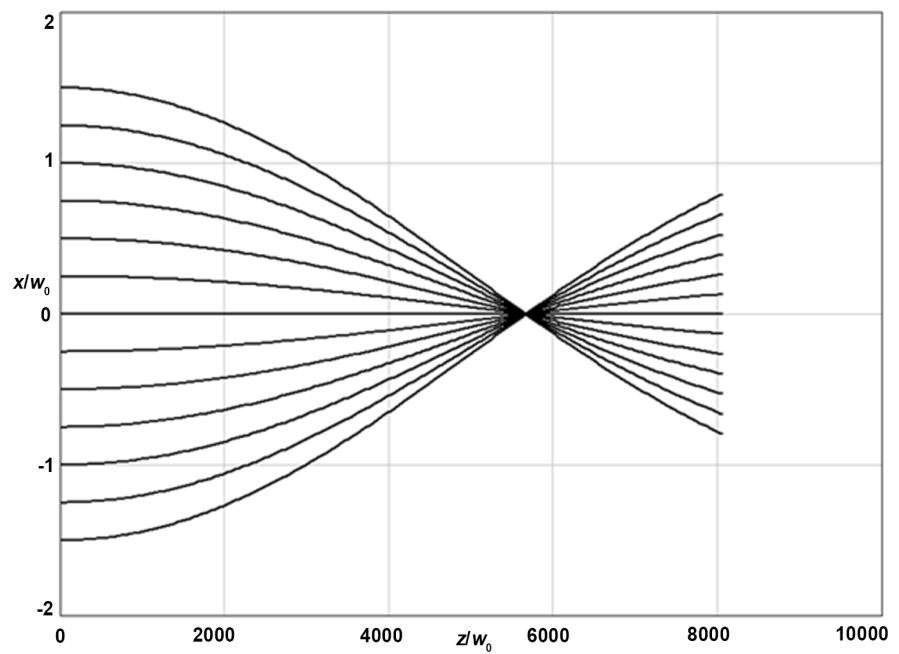

**Figure 7.** Lens-like potential: point-like focusing of a collimated beam in the absence of Wave Potential.

**Figure 9** shows, in its turn, the progressive sharpening of the beam intensity while reaching its finite waist.

A final application of Equations (20), clearly showing the diffractive role of the Wave Potential, is obtained by applying to a collimated Gaussian beam, launched along the $z$-direction, a potential of the form

$$V(z) = m\omega^2 z^2 / 2, \tag{33}$$

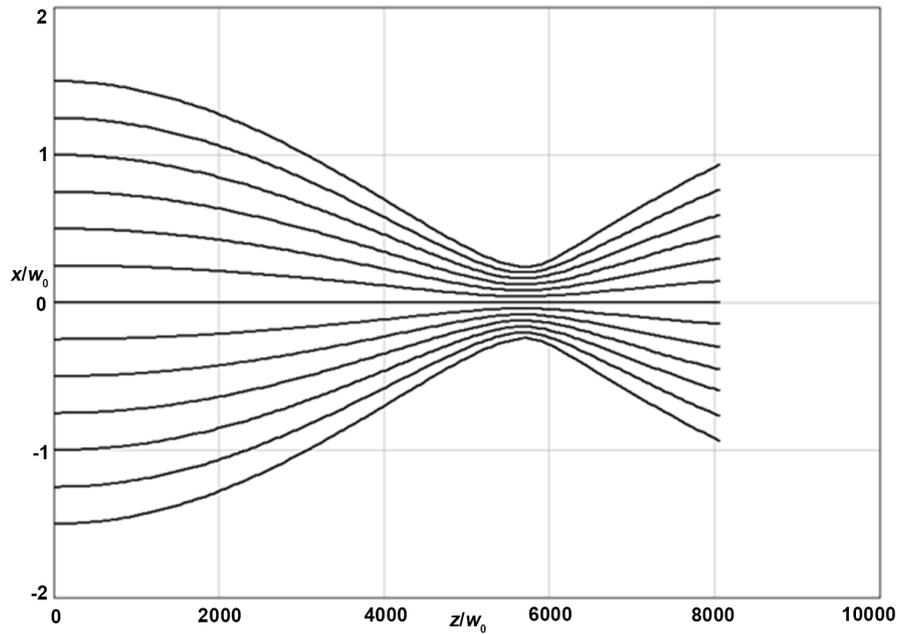

**Figure 8.** Lens-like potential: finite-focusing of a collimated beam in the presence of Wave Potential.

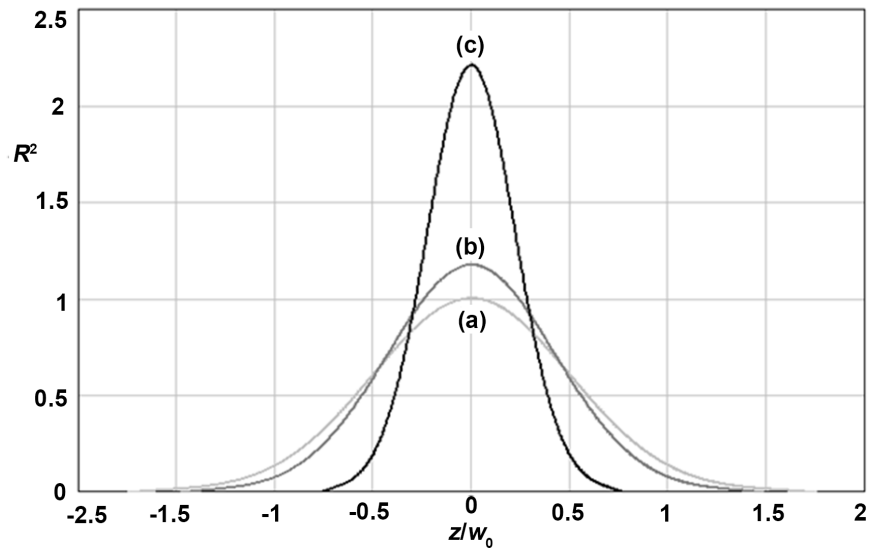

**Figure 9.** Lens-like potential: intensity-sharpening of a finite-focused beam in the presence of Wave Potential.

suggested by the classical case of particles harmonically oscillating around $z = 0$, with a period $T = 2\pi/\omega$, under an elastic force $-m\omega^2 z$. The particles moving along the beam axis are standard linear oscillators [8] [9], with well-known quantized energies

$$E_n = \left(n + 1/2\right)\hbar\omega, \tag{34}$$

which we assume to be shared by the whole beam. We show in **Figure 10** and **Figure 11**, respectively, the launch of the beam from $z = 0$ and its successive

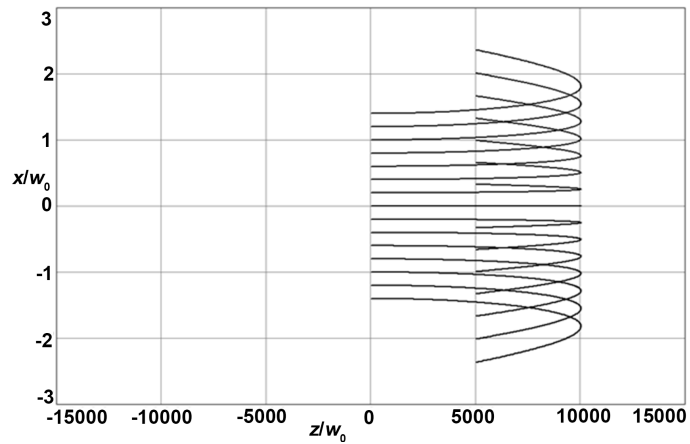

**Figure 10.** Harmonic potential: launching of the beam.

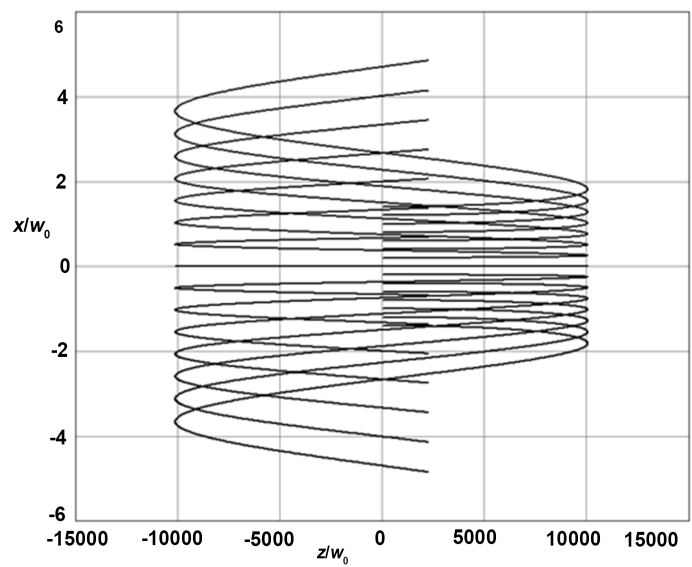

**Figure 11.** Harmonic potential: oscillations and diffractive widening of the beam.

oscillations around such a position. The beam is seen to oscillate between the positions

$$z_{\pm} \cong \pm\sqrt{2E_n/m\omega^2} \; , \tag{35}$$

while progressively widening under the diffractive action of the Wave Potential.

## 6. Born's Wave Function and Bohm's Approach

Even though, as we have seen, the wave-mechanical dynamics *in stationary external fields* is already adequately described by the *energy-dependent* Schrödinger Equation (17), one more interesting equation may be obtained from Equation (17) itself, making use of (12) and (13), in the form

$$i\hbar\frac{\partial\psi}{\partial t} = -\frac{\hbar^2}{2m}\nabla^2\psi + V(\mathbf{r})\psi \tag{36}$$

which is the standard *energy-independent* Schrödinger equation. Referring now, in order to fix ideas, to a discrete energy spectrum of Equation (17), and defining both the eigen-frequencies $\omega_n \equiv E_n / \hbar$ and the eigen-functions

$$\psi_n(\boldsymbol{r}, t) = u_n(\boldsymbol{r}) e^{-i\omega_n t} \equiv u_n(\boldsymbol{r}) e^{-iE_n t / \hbar}, \tag{37}$$

any linear superposition (with arbitrary constant coefficients $c_n$) of the form

$$\psi(\boldsymbol{r}, t) = \sum_n c_n \psi_n(\boldsymbol{r}, t), \tag{38}$$

turns out to be, as is well known [8] [9], a (deterministically evolving) solution of Equation (36). Such a function is a weighted average over the whole set of eigen-functions $\psi_n(\boldsymbol{r}, t)$, where the coefficients $c_n$ (in duly normalized form) may represent either a set of experimental results (in view of a statistical treatment) or an *ad hoc* mathematical assembling, in view of the construction of a particular "packet" of wave-trains. Each eigen-function is acted on by its own, energy preserving, Wave Potential $W(\boldsymbol{r}, E_n)$, and has therefore its own trajectories, provided by Equations (20). In Born's words [35], indeed, *"it's very attractive to interpret a particle of matter as a wave-packet due to the superposition of a number of wave trains. But this tentative interpretation comes up against insurmountable difficulties, since a wave packet of this kind is in general very soon dissipated"*.

Born himself proposed however, for the function (38), an interpretation [36] going much beyond a simple superposition, assuming it to represent the *most complete possible description* of the physical state of a particle whose energy is not determined, in the form of a simultaneous permanence (before observation) in its full set of eigenstates, according to the (duly normalized) probabilities $|c_n|^2$. Since Equation (36) is not, in itself, an ordinary-looking wave equation, its solution (38) is not, in itself, a wave. Such a mathematical object was called "*Born Wave-Function*", and became the pivot of Standard Quantum Mechanics (SQM), an intrinsically probabilistic theory which became the pillar of any further development of microphysics.

We find it useful to remind here a well known comment written by Jaynes [37] in 1989: "*Our present quantum mechanical formalism is an omelette that nobody has seen how to unscramble, and whose unscrambling is a prerequisite for any future advance in basic physical theory*". The pre-eminent role and the probabilistic nature assigned to the *energy-independent* Schrödinger equation (36)—scrambling all eigen-energies together—suggested, indeed, to extend the same probabilistic nature to the *energy-dependent* Equation (17), thus compromising the specific *guiding role* of the *energy-dependent* matter waves (14, 15) conceived by de Broglie and experimentally identified by Davisson and Germer.

An attempt to rescue this guiding role, while maintaining the properties of Born's Wave-Function, was performed in Bohm's approach [2] [3], where $\psi(\boldsymbol{r}, t)$ was written in the form

$$\psi\left(\mathbf{r},t\right)\equiv\sum_{n}c_{n}\psi_{n}\left(\mathbf{r},t\right)=R\left(\mathbf{r},t\right)\mathrm{e}^{iG\left(\mathbf{r},t\right)/\hbar} \qquad (39)$$

with real $R\left(\mathbf{r},t\right)$ and $G\left(\mathbf{r},t\right)$. The function $R^{2}\equiv\left|\psi\right|^{2}=\psi\psi^{*}$, in agreement with the SQM interpretation, was assumed to represent, in Bohm's words [2], "*the probability density for particles belonging to a statistical ensemble*". We made use here of the term $G\left(\mathbf{r},t\right)$, instead of Bohm's original term $S\left(\mathbf{r},t\right)$, *in order to avoid its instinctive a priori identification* with the Hamilton-Jacobi function of Equations.(14) and (15). Bohm's expression (39) was "shaped", in fact, on Equation (15) (*i.e.* on de Broglie's definite-energy matter waves, whose physical *reality* was established by the Davisson-Germer experiments) with the aim of depicting Born's "Wave-Function" as *a single pilot wave,* hopefully sharing and generalizing the nature and experimental evidence of de Broglie's guiding waves.

Since however, as we said before, the function $\psi\left(\mathbf{r},t\right)$ is not, in itself, the solution of a usual wave equation, let us look for its effective nature and relevance. By applying simple analysis to Equation (39) we get

$$\boldsymbol{\nabla}G\left(\mathbf{r},t\right)/m\equiv\frac{\hbar}{mi}\,\mathrm{Im}\!\left(\frac{\boldsymbol{\nabla}\psi}{\psi}\right)\equiv\frac{\hbar}{2mi}\frac{\psi^{*}\boldsymbol{\nabla}\psi-\psi\boldsymbol{\nabla}\psi^{*}}{\psi\psi^{*}}\,. \qquad (40)$$

Reminding therefore that the quantity $\mathbf{J}\equiv\frac{\hbar}{2mi}\left(\psi^{*}\boldsymbol{\nabla}\psi-\psi\boldsymbol{\nabla}\psi^{*}\right)$ represents, in SQM, a *probability current density* [8] [9], the term

$$\boldsymbol{\nabla}G\left(\mathbf{r},t\right)/m\equiv\mathbf{J}/R^{2} \qquad (41)$$

is seen to coincide with the velocity $\mathbf{v}_{prob}\left(\mathbf{r},t\right)$ at which "*the probability density is transported*" [5], in agreement *both* with Bohm's probabilistic assumption on the function $R^{2}$ *and* with the fluid-like equations (which we omit here for simplicity) obtained by Bohm [2] [3] by replacing Equation (39) into Equation (36) and separating real from imaginary parts.

## 7. Discussion

The basic equations of the *hydrodynamic* Bohmian theory, together with the ones of our *dynamical* system (20), may be summarized, respectively, in the following **Table 1** and **Table 2**. No time dependence is taken into account in the external potential $V\left(\mathbf{r}\right)$, since it wouldn't be strictly justified [38]. We remind however that, according to de Broglie [32], "*it is plausible to admit that, when V depends on time, the form of the equations be preserved as the general form of wave propagation in the Wave Mechanics of a single particle*".

The first equation of **Table 1** (to be time-integrated in parallel with the second one) is based on the *bold equalization* of the velocity of a (hopefully and permanently) narrow wave-packet centered at $\mathbf{r}$ with the *fluid-like* velocity $\mathbf{v}_{prob}\left(\mathbf{r},t\right)$, providing a (basically) hydrodynamic description which allows, to a certain extent, a pictorial visualization of SQM. This equalization represents however a handmade, external intervention mimicking, in agreement with our comment on Equation (39), the role of a "guidance velocity" *à la de Broglie*.

Table 1. Contains the equations of Bohm's *hydrodynamic* description.

$$\frac{\mathrm{d}\boldsymbol{r}}{\mathrm{d}t} = \frac{\hbar}{2mi} \frac{\psi^{*} \boldsymbol{\nabla} \psi - \psi \boldsymbol{\nabla} \psi^{*}}{|\psi|^{2}}$$

$$i\hbar \frac{\partial \psi}{\partial t} = -\frac{\hbar^{2}}{2m} \nabla^{2}\psi + V(\boldsymbol{r})\psi$$

Table 2. Contains, in its turn, the equations of the *dynamic* description proposed in the present paper.

$$\frac{\mathrm{d}\boldsymbol{r}}{\mathrm{d}t} = \frac{\boldsymbol{p}}{m}$$

$$\frac{\mathrm{d}\boldsymbol{p}}{\mathrm{d}t} = -\boldsymbol{\nabla}\big[V(\boldsymbol{r}) + W(\boldsymbol{r}, E)\big]$$

$$\boldsymbol{\nabla} \cdot (R^{2}\boldsymbol{p}) = 0$$

The equations of Table 1 are quite easily written, but not as much easily solved, and may give rise, in general, to analytical and computational troubles [4] [5]. We remind, in this connection, that the original formulation of the Bohmian theory involved an *energy-independent* "Quantum Potential" function $Q(\boldsymbol{r}, t) = -\frac{\hbar^{2}}{2m} \frac{\nabla^{2}\psi(\boldsymbol{r}, t)}{\psi(\boldsymbol{r}, t)}$, *formally similar* to (but *utterly different* from) our "Wave Potential" function $W(\boldsymbol{r}, E)$, whose (monochromatic) energy-dependence makes it apt to describe diffraction and/or interference processes. The exploitation of Bohm's Quantum Potential was faced, for instance, by means of iterative solutions leading to complex quantum trajectories [39], but didn't lead in general to easily practicable numerical procedures. The equation set of Table 1, bypassing the use of $Q(\boldsymbol{r}, t)$, represents therefore a more tractable approach.

The Hamiltonian system of Table 2, on the other hand, requiring no simultaneous solution of a Schrödinger equation and no bold conception, concerns the exact *dynamical* trajectories along which point-particles with an assigned total energy $E$ are driven by the dual mechanism of de Broglie's matter waves. Its numerical treatment—including $W(\boldsymbol{r}, E)$—is *not substantially less manageable than its classical counterpart*, as is shown by the examples of Section 5. Generalizing Kerner's comment [40] on the quantum solutions holding in the particular case of forced and/or damped oscillators, we may say that the particles, under the action of the Wave Potential, "*dance a wave-mechanical dance around their classical motion*".

## 8. Conclusions

In conclusion, the approach of the present paper is seen to provide a consistent *wave-mechanical* extension of Classical Dynamics. Starting from de Broglie's and Schrödinger's foundations of Wave Mechanics, we avoid (in agreement with Born's original caution alert [35] reported in Section 6) any wave-packet particle representation, and reach (together with a reasonably simple computability) a clear insight into the mechanism of wave-particle duality.

In the light of the emergence of "weak" experimental measurements, made possible by an increasingly powerful technology [41] [42] [43], a first peek is launched into Bohm's [2] [3] "hidden" variables.

## Conflicts of Interest

The authors declare no conflicts of interest regarding the publication of this paper.